\documentstyle[aps]{revtex}
\begin{document}
\draft
\title{Lamellae Stability in Confined Systems with Gravity}
\author{M. Bahiana, W. A. M. Morgado}
\address{Instituto de F\'\i sica, UFRJ\\
	Rio de Janeiro, RJ, Brazil}
\date{\today}
\maketitle
\begin{abstract}
The microphase separation of a diblock copolymer melt confined by hard walls 
and in the presence of a gravitational field is simulated by means of a cell dynamical system model.
It is found that the presence of hard walls normal to the gravitational field are key
 ingredients to the formation of well ordered lamellae in BCP melts. To this effect  the 
currents in the directions normal and parallel to the field are calculated along the 
interface of a lamellar domain, showing that the formation of lamellae parallel to the hard 
boundaries and normal to the field correspond to the stable configuration. 
Also, it is found thet the field increases the interface width.
\end{abstract}
\pacs{64.60.Cn,61.41.+e,64.75.+g}

Our main motivation for the present work has been to get insight about
the
mechanism responsible for the formation of well ordered layered samples
of
diblock copolymers, but, since the same striped pattern is observed in a
variety of systems (zebra skin, fingerprints, and the visual cortex, to
mention some), we expect  our results to be useful for the general
problem
of pattern formation in finite systems.
The formation of the striped pattern has been studied by several authors
and some factors important to
the phenomenon have already been identified: (i) hydrodynamic
interactions
combined with
the preferential wetting of a surface \cite{baoo}; (ii) annealing and
quenching of the structure, which are capable of relaxing the frustrated
structure locally \cite{chagun};(iii) long range interactions with an
external surface \cite{brown-chakra}.

We intend to study the influence of two different effects that have not
been included in the references quoted above (and 
still need to be better understood in the general context of block
copolymers): the simultaneous presence of the  gravitational field and
of a
rigid wall limiting the actual size of the sample, as in realistic
systems. For that matter we consider simulations of a block copolymer (BCP)
 melt based on a cell dynamical system
(CDS) model \cite{kitaoo} and the modified Cahn-Hilliard (CH) equation for 
driven spinodal decomposition \cite{yeung-dds}.
In both cases we analyze systems with a gravitational field normal to the hard
boundary with the intention of reproducing the presence of a substrate. 
As observed in real
systems, the layers are formed parallel to the substrate, even in the
absence of interaction with the wall. 

 Block copolymers are linear-chain molecules 
consisting of two 
each other. The subchains $a$ and $b$ are made of different monomer
units, $A$
 and $B$, respectively. Below some critical temperature $T_c$ these two
blocks
 tend to separate, but due to the covalent bond, they can segregate at
best 
locally to form a lamellar structure. The description of the microphase 
separation is similar to the spinodal decomposition of binary 
mixtures\cite{baoo,oosh}, so we borrow the CDS model proposed by
Kitahara 
{\it et al} for the spinodal decomposition with an external force field
with 
the appropriate addition of a term that makes the large uniform domains 
unstable. 
As usual, we assign a scalar variable $\psi(n,t)$ to each lattice site 
corresponding to the coarse-grained order parameter in the $n$-th cell
at time
 $t$ (time here is defined as the number of iterations). This order
parameter 
represents the difference $\psi_A-\psi_B$, where $\psi_A(\psi_B)$ is the
local
 number density of $A(B)$. The ingredients for the time evolution of
$\psi$ 
are: local dynamics dictated by a function with two symmetric hyperbolic 
attractive fixed points, diffusive coupling with neighbors,
stabilization of 
the homogeneous solution and conservation of $\psi$. The conservation,
when an
 external field is present, must be imposed by considering the Kawasaki 
exchange dynamics explicitly. The detailed explanation of this model is
found 
in \cite{kitaoo} for spinodal decomposition. With this, we come to final 
equation for a melt of even BCP molecules :
\begin{equation}
\psi(n,t+1)=(1-B)\psi(n,t)+
\left\langle\left\langle C\left(i,j;sgn [I(n,t)-I(j,t)]\right)
 (I(n,t)-I(j,t))\right\rangle\right\rangle,  \label{eq-model}
\end{equation}
where 
\begin{equation}
I(n,t) \equiv A \tanh\left(\psi(n,t)\right)-\psi(n,t)+
D\left(\langle\langle \psi(n,t)\rangle\rangle-\psi(n,t)\right)-hn_z
\end{equation}
is essentially the chemical potential.
$\langle\langle\star\rangle\rangle$
 is the isotropic space average of $\star$, $B$, $A$ and $D$ are
positive 
phenomenological constants. The parameter $B$ appears in this model to 
stabilize the solution $\psi=0$ in the bulk, for $B=0$ we have a model
for 
spinodal decomposition, in which the domains can grow without bound.
Scaling
 arguments have proved that $B\sim N^{-2}$, where $N$ is the
polymerization 
index (\cite{ooba1}). $h$ is the external field, which we assume is in
the $z$
 direction, and $n_z$ is the $z$ component of $n$. The collision
coefficient 
is given by: 
$C(i,j;\alpha)=[\psi_c+\alpha\psi(j)][\psi_c-\alpha\psi(i)]/\psi_c^2$,
where 
$\pm\psi_c$ are the fixed points of $A\tanh\psi -\psi$.

For all the simulations  we used a 128$\times$256 lattice, $A=1.2$,
$D=0.5$ 
and uniformly distributed random initial conditions. The external field,
when 
present, is parallel to the smaller dimension, which we call $z$. The
direction
 normal to the field will be called the $x$ direction. 
We impose no flux boundary conditions:$I(z+1)-I(z)|_{boundaries}= 0$.
Figure~\ref{fig-free} shows the pattern obtained
 after 20,000, 40,000 and 60,000 iterations, for B = 0.018 and h =
0.001. 

Domains initially aligned with the field present a varicose instability 
\cite{amalie} triggered by a density increase near the hard boundaries.
In this
varicose mode the bulges tend to grow making the coalescence of the
domains 
possible and we observe the formation of lamellae on the ``substrate''.
The
 central region, oriented along the field, shrinks with time. 
It is interesting to compare these patterns to the ones obtained by
Brown and
 Chakrabarti in \cite{brown-chakra}. These authors numerically
integrate the Cahn-Hilliard like equation of motion for the microphase
 separation dynamics with an extra term in the free energy functional
that
 takes into account a long-range interaction with a substrate.
 What they see is truly a
 surface effect, their patterns present  two regions: one with ordered 
lamellae, and another one with randomly oriented lamellae. That same
pattern
 had already been reported in
\cite{baoo} with a much simpler boundary condition: just keeping one
line
 with a fixed value of $\psi$. Here we observe patterns with  regions
where the
 lamellae are oriented 
normally to the field on top and bottom, ordered such that the denser
part is in contact with the substrate,
and another one in which the lamellae prefer to be
 parallel to the field, as found in systems with full periodic boundary conditions \cite{canela}. For the same boundary condition and $h = 0$,
 no layers are formed. To conclude the comparison, we have performed
simulations
 using a CDS model with the same form of surface field
as in
\cite{brown-chakra} and no bulk field to be compared to simulations with 
 the bulk field only.
 Adjusting the value of $h$ such that the boundaries in both cases
 have the same values of the field, we need a surface field one
 order of magnitude larger to obtain the same number of lamellae. The
patterns with the surface and bulk field are shown in
Figure~\ref{fig-surf} 

The presence of the field does not affect the linear stability about the 
homogeneous state $\psi=0$. We find that the solution $\psi=0$ is
unstable 
when $B \leq (A-1)^2/4D$, independently of the field. For the values $A
= 1.2$
 and $D = 0.5$ used in the simulations here, we should expect to observe 
pattern formation for $B \leq 0.1$. What we actually see from the
simulations
 is quite different, though. For example, for $B = 0.018$, 
$\psi = 0$ is stable for $h>0.02$. For smaller fields, there is pattern 
formation, but the amplitude of $\psi$ is smaller than that measured in
the
 $h = 0$ simulations. The decrease in amplitude is accompanied by an increase 
in the interface width, so the resulting pattern is less segregated than that 
without the field.
As for the domain morphology the early stages of
microphase separation are similar in
driven and
 non-driven systems \cite{canela}, in agreement with the linear stability analysis.

To understand the formation of lamellae on the hard boundaries we analyze the
stability of one vertical lamella in the presence of a gravitational field. 
It
is well known that a gravitational field enhances the stability of
interfaces parallel to itself\cite{yeung-dds} which is not observed near the 
hard boundary. To understand this we assume that close to the interface, 
the non-local term of the CH free energy for block copolymers is an 
irrelevant constant. In this spirit, we use the modified CH equation 
proposed in \cite{yeung-dds} for the spinodal decomposition in the 
presence of a gravitational field in the $z$-direction:
\begin{equation}
\frac{\partial \psi}{\partial t}=\nabla\cdot\left[ 
\left(1-\frac{\psi^2}{\psi_c}\right)\left(\nabla\frac{\delta F}{\delta \psi}-h\hat{z} \right)\right],
\label{CHfield}
\end{equation}
where $F[\psi]$ is the usual CH free energy, 
$F=\int d{\bf r}\left (\frac{1}{2}k(\nabla\psi)^2-\frac{1}{2}r_0\psi^2+\frac{1}{4}g\psi^4\right)$, for binary systems. 
The steady state solutions of (\ref{CHfield}) for a system with hard 
walls normal to the field satisfy
\begin{mathletters}
\label{currents}
\begin{equation}
j_x=-\partial_x(k\partial_x^2\psi+r_0\psi-g\psi^3)-k\partial_z^2\partial_x\psi=0
\label{jx}
\end{equation}
\begin{equation}
j_z=-\partial_z(k\partial_z^2\psi+r_0\psi-g\psi^3)-k\partial_x^2\partial_z\psi=h
\label{jz}
\end{equation}
\end{mathletters}

With this, we see that the presence of the
bottom wall forbids the  stable flux in the vertical 
direction so the system seeks the naturally
stable pattern, the one parallel to the wall. From (\ref{jz}) it can be 
shown, for the case of horizontal lamellae, that the interface becomes less
 sharp with increasing $h$. This effect is noticeable in 
Fig.~\ref{fig-surf}: all the patterns where plotted with the same grey 
scale so it is clear that pattern (b), with gravitational field, is more 
diffuse. This result
 can be understood if we assume that 
$\psi(z)=\psi_e\tanh(z/\xi)+\epsilon\psi_1+O(h^2)$ for a particular 
interface located at $z=0$, where $\psi_e=\sqrt{r_0/g}$, 
$\xi=\sqrt{2k/r_0}$ and $\epsilon=h/k$. We obtain $\psi_1= -
\frac{z}{4\xi}\psi_e$, hence
\begin{equation}
\psi (z) = \psi_e\left(\tanh\frac{z}{\xi}-\epsilon\frac{z}{4\xi}\right),
\end{equation}
and
\begin{eqnarray}
\psi'_{h=0}(0)&=&\frac{\psi_e}{\xi}\nonumber\\
\psi'_{h\neq 0}(0)&=&\frac{\psi_e}{\xi}\left(1-\frac{\epsilon}{4}\right)
<\frac{\psi_e}{\xi},
\end{eqnarray}
which shows that the interface has an elongated profile in the presence of 
the field.
In summary, we  conclude that the presence of hard walls normal to the 
gravitational field are key ingredients to the formation of well ordered 
lamellae in BCP melts. Other interactions like hydrodynamics and preferential 
wetting of the substrate may enhance this effect, but we find that the bulk 
field considered here is more efficient than a
surface field for ordering the lamellae, and involves a different mechanism 
of lamellae formation. We also find that the gravitational field increases 
the interface width, producing less segrgated patterns.

\section*{Acknowledgments}
This work was partially supported by CNPq (Brazil) and Faperj (state
of Rio de Janeiro, Brazil). The simulations were done 
on the Cray-J90 computer of the NACAD, COPPE/UFRJ, Brazil.

\begin{figure}
\caption{Simulation of a driven system with no flux boundary conditions.
Here, 
$A = 1.2$, $B = 0.018$ and $h = 0.001$ after (a) 5,000, (b) 20,000 and
(c) 60,000 iterations. Lamellae oriented normal to the external field 
accumulate at
the hard boundaries, as observed in real systems.}
\label{fig-free} 
\end{figure}
\begin{figure}
\caption{Here we compare the long-range surface interaction of the form $h_s/z^3$ as proposed by Brown and Chakrabarti. For all three patterns $B = 0.018$. 
In (a) is the pattern obtained with surface interaction only and $h_s$ = 0.001. (b) corresponds to the pattern with gravitational field $h$ = 0.001, and (c) 
surface field $h_s$= 0.01. For the values chosen, (a) and (b) have the 
same field at $z$ = 1, but the pattern with gravity is more ordered. 
Close to the substrate patterns (b) and (c) present the same number of l
amellae but the filed in (c) is ten times larger, showing that the bulk 
field is more effective in producing well ordered lamellae.}
\label{fig-surf}
\end{figure}
\end{document}